\documentstyle[prl,aps,twocolumn,epsfig]{revtex}    
\begin{document}    

\draft   

\twocolumn[ 
\hsize\textwidth\columnwidth\hsize\csname@twocolumnfalse\endcsname 

\title{Renormalization Group Study of the Intrinsic Finite Size Effect in 
2D Superconductors}    

\author{Stephen W.~Pierson$^{1,*}$ and Oriol T.~Valls$^{2,\dag}$}    
\address{$^1$Department of Physics, Worcester Polytechnic Institute, 
Worcester, MA 01609-2280}    
\address{$^2$School of Physics and Astronomy and
Minnesota Supercomputer Institute \\  University of Minnesota, Minneapolis, 
MN 55455-0149}  

\date{\today}    
\maketitle    
\begin{abstract}    
Vortices in a  thin-film superconductor interact logarithmically out 
to a distance on the order of the two-dimensional (2D) magnetic penetration 
depth $\lambda_\perp$, at which point the interaction approaches a constant. 
Thus, because of the finite $\lambda_\perp$, the system exhibits what
amounts to an
{\it intrinsic} finite size effect. It 
is not described by the 
2D Coulomb gas but rather by the 2D Yukawa gas (2DYG).  To study the 
critical behavior of the 2DYG, we 
map the 2DYG to the massive sine-Gordon model and then perform a 
renormalization group study to derive the recursion relations and to
verify that $\lambda_\perp$ is a relevant parameter.  We solve the recursion 
relations to study important physical quantities for this system including 
the renormalized stiffness constant and the correlation 
length. We also address the effect of current on this system to explain why
finite size effects are not more prevalent in experiments given that the
2D magnetic penetration depth is a relevant parameter.
\end{abstract}    
] 

\narrowtext    
\section{Introduction}    
\label{sec:intro} 
Interest in the critical behavior of vortices in two-dimensional 
(2D) superconductors can be traced back to at least the studies of 
Kosterlitz and Thouless\cite{kt73,k74} and Berezinskii\cite{b71} in
the early 1970's. It was stated in Ref.~\onlinecite{kt73} however,
that the theory as formulated did not apply to 2D superconductors 
where there is a coupling
to the magnetic field. Because of this coupling, the interaction is 
logarithmic only up to the length scale $\lambda_\perp=2 \lambda^2/d$, 
(where $\lambda$ is the London penetration depth and $d$ is the film 
thickness,) at which point it approaches a constant.\cite{pearl64} 
The application of the Kosterlitz-Thouless-Berezinskii (KTB) transition 
to 2D superconducting systems dates only to the late 
1970's and the work of Beasley, Mooij, and Orlando\cite{beasley79} and 
of Halperin and Nelson\cite{hn79} who pointed out that $\lambda_\perp$ can be quite
large (relative to the film size) in superconducting thin films and that therefore 
these films should behave as the charged analog of superfluid thin films.
As a result, a flurry of experimental papers reported investigations of KTB 
behavior in thin-film superconductors in the early 
1980's.\cite{fiory83,kadin83}

Sparked partially by the discovery of the high-temperature 
superconductors, interest in the the critical behavior of 2D 
superconductors (SC's) and Josephson Junction arrays (JJA's) has been renewed 
in this decade. Sujani, Chattopadhyay, and Shenoy\cite{shenoy94} have 
derived recursion relations for 2D superconductors that incorporate 
nonzero current drives. Minnhagen and coworkers\cite{jonsson97} have been 
investigating the dynamics of the transition numerically while Ammirata 
{\it et al.}\cite{ammirata99,pierson99} have applied dynamic scaling 
to $I$-$V$ data from 2D superconductors and JJA's.  

The limiting large scale $\lambda_\perp$, a length intrinsic to 2D 
superconductors, acts as an unavoidable source of finite size
effects. Other sources, characterized by other lengths, are 
extrinsic, arising from the finite size (or extent) of a system and 
from a current. Some of the recent work on 2D SC's and JJA's has focused 
on the expected finite size effects (of any origin) in these systems. 
Simkin and Kosterlitz\cite{simkin97} have investigated numerically and 
analytically the finite size effects in these systems, while Repaci 
{\it et al.}\cite{repaci96} claimed experimental evidence of finite size 
effects in ultra thin film YBCO. Herbert {\it et al.}\cite{herbert98} 
have extended this work to JJA's. We will focus here on the influence 
of a finite $\lambda_\perp$, which we refer to as the {\it intrinsic}
finite size effect.

Despite all of the work on 2D SC's and JJA's, there is still no analytical 
theory of the influence of intrinsic finite size effect on their critical 
properties. This is particularly striking since it is generally accepted 
that the intrinsic finite size effect should be relevant, yet reports of
its experimental observation are relatively sparse. Our objective in this 
paper is to fill in the theoretical gap and to address the small number of 
experimental reports of finite size effects. We perform a RG study of the 
vortices in 2D superconductors by modeling them as a 2D Yukawa gas (2DYG) 
and then mapping this system to the massive sine-Gordon (SG) Hamiltonian. 
We then perform a momentum-space shell RG study  to derive the recursion
relations, which we then analyze to study the shift 
in the critical temperature and the behavior of the system near it. 
We will also address the effect of the presence of a driving current
on the critical behavior of the system.

This paper is organized as follows. In Section \ref{sec:model}, 
the partition function for the 2DYG is given and shown to map 
to the massive sine-Gordon. In Section \ref{sec:rg}, the
recursion relations are derived. These equations are analyzed 
in Section \ref{sec:analysis} and their experimental ramifications
are explored in Section \ref{sec:exp}. Finally, the paper is
summarized in Section \ref{sec:summ}.

\section{Model and Renormalization Group Study} 
\label{sec:model} 

As indicated above, our approach is to model the vortices in two dimensional 
superconductors as a vortex gas interacting via the 2D 
Yukawa potential. Therefore, we begin this section by describing the 
partition function for the 2DYG. We then will map it to the massive 
sine-Gordon model, and  perform
a RG study on the latter.

\subsection{Model}

The partition function for the neutral vortex gas ($N_+$ 
vortices and $N_-=N_+$ 
antivortices) in a 2D superconductor is
\begin{eqnarray}
Z= && \sum_N y^{2N} {\frac{1}{(N!)^2}}
\int d^2r_1\int d^2r_2...\int d^2r_{2N} 
\nonumber \\
&&\times \exp\left[-{\frac{\beta}{2}}\sum_{i\not= j}p_i p_j V(\left| 
{\bf r}_i-{\bf r}_j \right|)\right]
\label{partfunc}
\end{eqnarray}
where $2N$ is the total number of
particles ($N=N_{\pm}$), ${\bf r}_i$ are the coordinates 
of the $i$th charge $p_i=\pm p$ and $y=\exp(\beta\mu)$ 
is the fugacity, where
$\mu=- E_c$ and $\beta=1/k_BT$. 

$V(R)$ is the interaction 
of a pair of vortices, the strength of which is $p^2=\pi 
n^{2D}_s\hbar^2/2m$ ($n^{2D}_s=n_sd$ is the areal superfluid 
density, $n_s$ is the superfluid electron density, and $m$ is the mass 
of a free electron.) For the 2D Coulomb gas, the interaction is 
logarithmic $V(R)= \ln (R/\tau)$ (where $\tau$ is the vortex core 
size) for all lengths. But for a 
superconductor, the interaction is only logarithmic out to
$\lambda_\perp$,\cite{pearl64} at which point it approaches 
a constant as $1/R$ due to the coupling to the magnetic field. 
We will approximate this behavior for the potential by writing
it in terms of a 
modified Bessel function,
\begin{equation}
V(R)=\ln(\lambda_\perp/\tau)-{\rm K}_0(R/\lambda_\perp),
\label{potential}
\end{equation}
which has the limits calculated by Pearl.\cite{pearl64} 
We will refer to the vortex gas in 2D superconductors as 
the 2D Yukawa Gas (2DYG) since this potential is the 2D 
equivalent of the more familiar 3D Yukawa potential.

\subsection{Mapping to the Massive sine-Gordon}

A standard way of treating the 2DCG is to map it to the 2D 
sine-Gordon  Hamiltonian.\cite{knops80}
For the case of the 2D vortex gas in a superconductor, as we now
review, the correct mapping is to the {\it massive} 
sine-Gordon model:\cite{nicolo83}
\begin{equation}
-{\cal I}\int d^2r \left\{\left[{\bf\nabla} \phi({\bf r})\right]^2
+\left[m\phi({\bf r})\right]^2\right\} +2y\int d^2r \cos \phi({\bf r}) 
\label{massiveSG}
\end{equation}
where ${\cal I}=1/[2\pi\beta p^2]$ is the coupling strength 
or ``stiffness,'' $\phi$ is the SG field and $m$ is the mass, 
which we will relate to $\lambda_\perp$ below. 
For $m=0$, Eq.~(\ref{massiveSG}) reduces to the SG Hamiltonian.

In Ref.~\onlinecite{pierson94} the mapping of a layered vortex gas 
with arbitrary interactions 
to a SG-like Hamiltonian was studied. 
By examining the 2D limit of this mapping, we can establish the
link between the 2DYG and the massive sine-Gordon. Following 
Ref.~\onlinecite{pierson94}, one
can show that Eq.~(\ref{partfunc}) can be written as 
\begin{equation}
Z={\frac {\int D\phi({\bf r})  \exp\left\{-H_o+2y\int d^2r
\cos[\phi({\bf r})]\right\} }
{\int D\phi({\bf r}) \exp[-H_o]}},
\label{zsg}
\end{equation}
where the Hamiltonian $H_o$ is  Gaussian 
in the field $\phi$. The connection between the sine-Gordon 
field and the potential $V(R)$ is given by:
\begin{equation}
\beta p^2 V(R)=<\phi({\bf R})\phi(0)>_0-<\phi^2(0)>_0,
\label{connection}
\end{equation}
where $\langle...\rangle_{0}$ denotes an average with
respect to a Hamiltonian $H_o$ Gaussian in the field $\phi$.

Making use of Eq.~(\ref{connection}), one can readily show 
that $V(R)=\ln(\lambda_\perp/\tau)-{\rm K}_0(R/\lambda_\perp)$ when 
$H_0={\cal I}\int d^2r \{[{\bf\nabla} \phi({\bf r})]^2
+[m\phi({\bf r})]^2\}$, $\beta p^2=1/[2\pi {\cal I}]$, 
and $m=\tau/\lambda_\perp$. 
The calculation of the right hand side of Eq.~(\ref{connection}) 
is straightforward 
and quite similar to that
done for the layered case in Ref.~\onlinecite{pierson92}. 
(See the calculation of $g_s(R)$ in Appendix B of that 
reference.) After performing the Gaussian integral over the 
fields, one is left with
\begin{equation}
\beta p^2 V(R)=\int {\frac{d^2q}{(2\pi)^2}}{\frac{1-
\cos({\bf q}\cdot{\bf R})}{2{\cal I}(q^2+m^2)}},
\label{intermediate}
\end{equation}
where the angle is integrated from $0$ to $2\pi$ and $q$ is 
integrated from zero out to its infrared cutoff $q_0\sim2\pi/\tau$.
Performing the integral over the angle first and then over $q$, 
one obtains
\begin{eqnarray}
\beta p^2V(R)={\frac{1}{2\pi{\cal I}}}
&& \Bigl\{[1-{\rm J}_0(q_0R)]\ln R
\nonumber \\
&& -{\rm K}_0(mR)-\ln (mR)\Bigr\},
\label{final}
\end{eqnarray}
where $R$ is now expressed in units of $\tau$ and $q$ in units of
$1/\tau$. Neglecting the oscillating behavior\cite{ma}
 of the zeroeth order Bessel function 
${\rm J}_0(q_0R)$ at large $R$, which arises from the sharp
cutoff, this
has the desired behavior of Eq.~(\ref{potential}).

This completes that mapping between the 2DYG and the massive
2D SG Hamiltonian. We will now proceed to a RG study on the latter.

\subsection{Renormalization Group Study}
\label{sec:rg}

The advantage of the above mapping is that it is easier to perform 
a momentum space RG study on Eq.~(\ref{zsg}) than a real space RG 
calculation based on  Eq.~(\ref{partfunc}). 
In this subsection we show how to perform a RG study
in momentum space on Eq.~(\ref{zsg}) to obtain the 
recursion relations for the three parameters in the 
system ${\cal I}$, $y$, and $m$. The recursion relations 
for this system were previously derived in 
Ref.~\onlinecite{ichinose94}, but as we will explain 
below, a technical point was neglected which rendered the results
formally incorrect. 

To briefly overview the process, the RG 
study consists of integrating out the large wavevector 
$q$ components in a thin shell: $[q_0(1-\epsilon)<q<q_0]$, 
which correspond to small scale structure. The system is 
then shrunk back so that the cutoff in momentum space is 
restored to the original. We follow the procedure of Knops 
and den Ouden\cite{knops80} as outlined in 
Ref.~\onlinecite{pierson92} for a layered vortex gas. 
Because the calculation is closely related to the previous 
work of the present authors,\cite{pierson92} we do not need to show 
much detail here. 

The first step in the RG study is to Fourier 
analyze the first term in the free energy,
\begin{eqnarray}
F=-{\frac{1}{2}} && \int {\frac{d^2q}{(2\pi)^2}}\phi(q)
\phi(-q) [{\cal I} (q^2 + m^2)] 
\nonumber \\
&& + {2 y}  \int d^2r \cos [\phi({\bf r})],
\label{freeenergy}
\end{eqnarray}
it is then convenient to define the Gaussian terms in 
Eq.~(\ref{freeenergy}) as
\begin{equation}
F_0=\phi(q)\phi(-q) [I (q^2 + m^2)]. 
\label{f0}
\end{equation}

There are two differences between the 2DYG and the layered SG
model of 
Ref.~\onlinecite{pierson92}: First,  $F_0$ for the layered case is not
given by  Eq.~(\ref{f0}), but rather by
$\phi_k(q) \phi_{-k}(-q) [ {\cal I} q^2 + 2 K(1-\cos (ks))]$ 
where $s$ is the interlayer distance and $k$ is the $z$-axis
wave vector. ($K$ also has a different meaning there.) 
Second, the system here is purely two-dimensional and so the 
integration over the third dimension $k$ in Ref.~\onlinecite{pierson92} 
does not appear. Keeping these two distinctions in mind and carrying 
out the RG 
study in the manner of Ref.~\onlinecite{pierson92}, one finds
\begin{equation}
{\cal I}'={\cal I} +\epsilon A({\cal I},m),
\label{rrI}
\end{equation}
\begin{equation}
y'=y\{1+\epsilon[2-f(R=0)/2]\},
\label{rry}
\end{equation}
and
\begin{equation}
m'=m\{1+\epsilon[1-y^2A({\cal I},m)/{\cal I}]\},
\label{rrm}
\end{equation}
where 
\begin{equation}
A({\cal I},m)=\int d^2R ({\bf q} \cdot {\bf  R})^2
f(R) e^{-g(R)/2},
\label{AIm}
\end{equation}
\begin{equation}
f_s(R)=\int_0^{2\pi}\int_{q_0(1-\epsilon)}^{q_0} 
{\frac{d^2q}{(2\pi)^2}}  {\frac{e^{i{\bf q}
\cdot {\bf R}}}{{\cal I}(q^2+m^2)}}, 
\label{fR}
\end{equation}
and
\begin{equation}
g_s(R)=\int_0^{2\pi}\int_0^{q_0}
{\frac{d^2q}{(2\pi)^2}} {\frac{1-\cos ({\bf q} \cdot
{\bf R})}{{\cal I}(q^2+m^2)}}.
\label{gR}
\end{equation}
Integrating Eqs.~(\ref{fR}) and (\ref{gR}) we have
\begin{equation}
f(R)={\frac{\epsilon}{2\pi{\cal I}(1+m^2)}}J_0(R),
\label{fR2}
\end{equation}
and
\begin{equation}
g(R)={\frac{1}{\pi{\cal I}}}
\left[(1-{\rm J}_0(q_0R))\ln R-{\rm K}_0(mR)-\ln (mR)\right].
\label{gR2}
\end{equation}
Substituting these into $A({\cal I},m)$ and expanding to 
first order in $m$, one finds\cite{pierson92}
\begin{equation}
A({\cal I},m)=A({\cal I}) - A_m({\cal I})m\ln m.
\label{AcalI}
\end{equation}
The values of $A({\cal I})$ and $A_m({\cal I})$ depend upon 
the cutoff used and 
the nature of core used. Here, as in Ref.~\onlinecite{pierson92}, 
we will assume 
a sharp cutoff in momentum space and a hard-core. 

The recursion relations can be put in more
convenient differential form. Introducing also
the stiffness constant, $K=1/[2\pi{\cal I}] (\equiv\beta p^2)$, we can
write the final result in the form
\begin{equation}
{\frac{dK}{d\epsilon}}=-{\frac{1}{2}}K^2y^2(1-4.5m\ln m),
\label{rra}
\end{equation}
\begin{equation}
{\frac{dy}{d\epsilon}}={\frac{y}{2}}(4-{\frac{K}{1+m^2}}),
\label{rrb}
\end{equation}
and
\begin{equation}
{\frac{dm}{d\epsilon}}=m(1-y^2{\frac {K}{4}}).
\label{rrc}
\end{equation}
Our definition of the parameter $K$ differs from others in the
literature by a factor of $2\pi$.

The $m=0$ limit of these equations correctly 
reduce to the original Kosterlitz recursion relations for 
the 2DCG.\cite{k74} Furthermore, the $m$ corrections to those 
relations are of the form that one would expect 
physically.\cite{pierson94b} For example, in the recursion 
relation for $K$, the screening effect of small pairs on $K$ 
is increased by $m$ (since $m\ll 1$). This is because the 
vortex pairs are more weakly bound for a superconductor 
than for a superfluid and so they are more susceptible to screening. 
(This is quite consistent  with the opposite result found
in the case of a layered vortex 
gas. There, the vortex pairs 
are more {\it strongly} bound than in the 2DCG and the 
screening effects in the recursion relation for $K$ 
{\it diminished}.\cite{pierson94b,friesen95a}) In 
the recursion relation for $y$, the number of vortex pairs  
increases in the presence of finite size effects because the 
energy of a vortex pair is reduced. In the recursion relation 
for $m$, there are two terms. The first enters through the 
shrinking scale step of the RG calculation. 
The second, of $O(y^2)$, enters through 
the integrating out of small scale structure, which in this case 
consists of the small vortex pairs. This explains the  $y^2$ dependence. 
This second term arises in a somewhat complicated way from  
the renormalization of the coefficient of the $({\bf\nabla} 
\phi({\bf r}))^2$ term in Eq.~(\ref{massiveSG}). This enters into 
the mass term $m$ because the coefficient of  
$(m \phi({\bf r}))^2$ is ${\cal I}m^2$ and so $m$ must be 
renormalized in the second term to compensate the 
renormalization of ${\cal I}$ in the first term.

The recursion relations Eqs.~(\ref{rra})-(\ref{rrc}) differ 
from those of Ref.~\onlinecite{ichinose94} in the $m$ correction 
to the equation for $K$. This is because of an 
approximation made there which was pointed out to be 
incorrect by Knops and den Ouden.\cite{knops80} The 
approximation, common in the literature,\cite{wiegmann78,ohta78} 
is to Taylor expand a cosine term (see the first term
in Eq.~(A8) of Ref.~\onlinecite{pierson92}) and then to 
neglect the higher order terms. As Knops and den Ouden 
showed, the higher order terms are not negligible and 
in fact contribute in a significant way. In our case, 
the contribution of the higher order terms is to give 
an $m$ correction to the recursion relation for $K$ 
that is opposite in sign to that obtained by Ichinose 
and Mukaida.\cite{ichinose94} (As discussed in the 
previous paragraph, the sign found here makes more 
physical sense.) However,
the $m$ correction to the recursion relation  that most 
influences the system and the integration of the 
recursion relations turns out to be that for $y$,
Eq.~(\ref{rrb}), and this renders the effect of the error
discussed above less harmful.

\section{Analysis of the Recursion Relations}    
\label{sec:analysis}  

By integrating Eqs.~(\ref{rra})-(\ref{rrc}), the analysis 
of the effects of the finite length scale on the 2DCG is 
possible. One sees at once that the size effect, or mass,
term is relevant. Because $y^2$ is small, the growth of $m$ with 
$\epsilon$ is nearly exponential, which leads to the 
intrinsic finite size effects dominating the integration of the 
recursion relations at large length scales and, in certain 
temperature regimes, at smaller length scales too. In this 
section we analyze this behavior and the temperature 
regimes where these effects are most significant. 
We will first analyze the RG flows, move on to discuss the 
correlation length, and then examine the temperature dependences 
and the behavior of the renormalized stiffness constant.

\begin{figure}
\centerline{
\epsfig{file=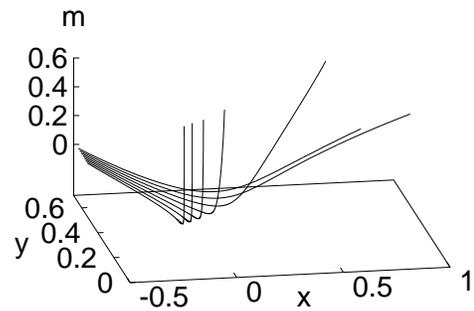,width=3.4in}}
\caption{The RG flows for the 2DYG. Because $\lambda_\perp$ is a relevant
parameter, all flows will eventually flow out of the $x$-$y$ plane. The
direction of the flows in the figures can be inferred from the
location of the starting points, as given in the text.}
\label{rgflows}
\end{figure}

In Figure \ref{rgflows}, we show a plot the ``flows'' of 
the three system parameters as a function of length scale. To
facilitate the comparison of these results with those of the 2DCG in
the literature, we use the parameter $x=4/K-1$.  The flows are 
obtained by integrating the recursion relations for representative 
initial values of the three parameters: $K_i=8$ ($x_i=-0.5$), 
$m_i=10^{-5}$, and $y_i=0.56$, 0.58, 0.60, 0.62, 0.64, 0.66, 0.68. (The 
initial values of each parameter will be denoted below by the subscript 
$i$.) The recursion relations are integrated until they cease to be valid, 
which is when $y$ or $m$ become large [$y(\epsilon)\ge 0.7$ or 
$m(\epsilon)\ge 0.7$] and the approximations that we made in deriving them 
break down. As one can see, the flows always tend towards large $m$ 
at the end of the flows. This reflects that $m$ ($\lambda_\perp$)
is a relevant parameter: finite size effects will ultimately
influence or determine the critical behavior. This parameter dominates
the end of the flows, which corresponds to the largest length scales: 
finite size effects result in free vortices, and the average distance
between these free vortices corresponds to these long lengths. This 
will become more evident as we examine other aspects of the flow.

\begin{figure}
\centerline{
\epsfig{file=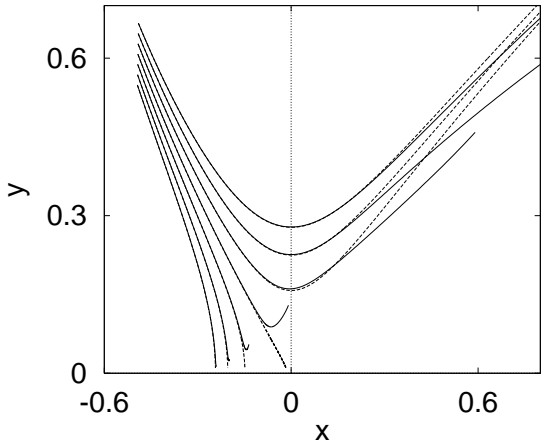,width=3.4in}}
\caption{The projection of the RG flows for the 2DYG (solid lines) 
onto the $x$-$y$ plane. For comparison, the RG flows of the 2DCG (dashed
lines) are also included.}
\label{projection}
\end{figure}

A projection of these flows  onto the $x$-$y$ plane 
is shown (solid lines) in Figure \ref{projection} to better 
illustrate the effect 
of $m$ on the 2DCG. Also plotted in the figure are the RG 
flows (dashed lines) for the 2DCG (i.e., the $m=0$ 2DYG). We first explain the
RG flows for the 2DCG. The low temperature ($T<T_{KTB}$) flows 
all iterate to $y=0$, which corresponds to zero vortex density since 
there are no pairs at the largest separations for $T<T_{KTB}$. 
This follows from the relation between vortex fugacity and vortex density.
The high temperature ($T>T_{KTB}$) flows tend toward $y=0$ initially 
but ultimately towards $y=\infty$.  The initial decrease reflects 
the vortex {\it pair} density decrease, while the upturn and subsequent 
rise is a result of the increasing number of {\it free} vortices above the
transition temperature. The isotherm that divides the 
low temperature flows from the high temperature flows represents the 
critical temperature. For the linearized recursion relations
for the 2DCG, the critical isotherm is the line $y=-x$.

Looking now at the RG flows of the 2DYG contrasted to the 2DCG,
one sees that it is at the largest values of $\epsilon$ 
that the largest deviation from the 2DCG flows occurs. What 
is also visible here and in Figure \ref{rgflows} is that none of 
the flows end at $y=0$ but all eventually will tend towards 
$m,y=\infty$. In other words, all of the flows have the same 
qualitative behavior. This confirms the expected behavior that 
there is no phase transition in the 2DYG since there are free 
vortices present at all temperatures. Further, the fact that 
more RG flows move towards the higher temperature limit than 
for the 2DCG shows that the apparent phase 
transition temperature does shift toward lower temperature 
before being wiped out. Along with the tendency of the 2DYG
flows to move toward $y=\infty$, we note that the corresponding
tendency for the flows to more rapidly approach large $x$ 
than those of the 2DCG, which is especially evident for the 
three highest temperatures in Figure \ref{projection}. This 
reflects the tendency of the renormalized stiffness constant
$K$ to approach zero more rapidly in the 2DYG.

\begin{figure}
\centerline{
\epsfig{file=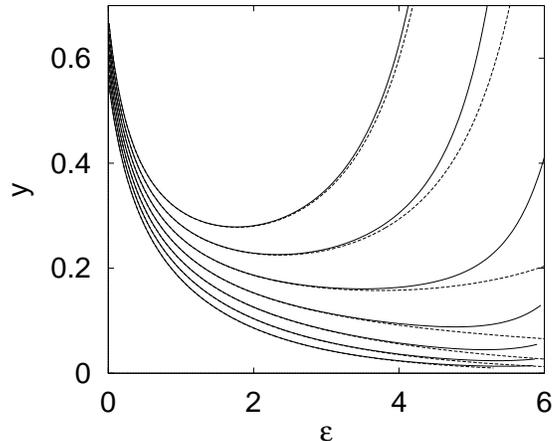,width=3.4in}}
\caption{The vortex fugacity $y$ as a function of $\epsilon$, 
the logarithm of the length scale for the 2DCG (dashed lines) and
the 2DYG (solid lines) with $m_i=10^{-5}$. The initial fall is due to 
the decreasing number of vortex pairs as a function of increasing size. 
The increase in $y$ at larger length scales represents free vortices, 
which are always present in the 2DYG.}
\label{ye}
\end{figure}

It is instructive to look solely at $y$ as a function of 
length scale because this quantity represents the density 
of vortices as a function of length scale. For the same 
temperatures as in Figures \ref{rgflows} and \ref{projection}, 
we plot $y$ as a function of $\epsilon=\ln \tau$ in Figure 
\ref{ye} (solid lines) along with $y(\epsilon)$ for the 2DCG
(dashed lines). 
For temperatures far below $T_{KTB}$, one can see that the 
density of vortices is dropping rapidly with length scale. As 
mentioned above this range 
of $\epsilon$ represents vortex pairs. At large enough length scales, 
there is always an upturn in $y(\epsilon)$. This is due to free 
vortices present because of finite size effects.  As the 
temperature is increased, while still remaining below $T_{KTB}$, 
the upturn in $y$ at large $\epsilon$ moves to lower 
values of $\epsilon$. At temperatures above $T_{KTB}$, 
the behavior is similar but more pronounced. This is 
because the upturn in $y$ at large $\epsilon$ is always 
present in the 2DCG for $T>T_{KTB}$ since free vortices 
can be spontaneously created in this temperature range. 

Further insights can be gained through an examination of 
the correlation length, which is obtained as follows. As 
one integrates the recursion relations, $y$, $K$, and $m$ are
parameterized by $\epsilon$, the natural log of the length scale.
As mentioned above, the recursion relations are integrated until 
they break down. The value of $\epsilon$ at that point is 
denoted by $\epsilon_{max}$. Because there is only one length scale 
in the system and it is the vortex correlation length 
$\xi(T)$, one makes the association $\epsilon_{max}=\ln 
\xi(T)$.\cite{k74,pierson95a,pierson95b} The temperature dependence 
enters through the first integral of the recursion relations of the 
(non-linearized) 2DCG: $y_i^2-2x_i^2+\ln(1+c_i)=c$ where $c\propto 
T-T_{KTB}$. For small $m$, we assume that this association remains valid.

The meaning of the 2DCG correlation length below $T_{KTB}$, 
$\xi_-(T)$, deserves mention. Kosterlitz\cite{k74} originally 
defined $\xi_-(T)$ to be infinite because the susceptibility 
below the transition temperature is infinite. Ambegaokar 
{\it et al.},\cite{ahns78} on the other hand, 
defined a finite diverging correlation length for 
$T<T_{KTB}$, based on the critical behavior of the 
dielectric constant. (Because the two descriptions have different 
meanings, they do not contradict one another.) Later, Simkin
and Kosterlitz\cite{simkin97} stated that $\xi_-(T)$ is the length 
scale at which $K(\epsilon)$ is essentially at its
asymptotic value\cite{simkin97}. In other words, $\xi_-(T)$ has
meaning both above and below $T_{KTB}$ and here, we take it to
represent the characteristic size of the largest vortex pairs. $\xi_+(T)$,
on the other hand, is defined by where the recursion relations 
become invalid and can be thought of as the length scale at which
the influence of free vortices is significant. The relationship 
of $\xi_-$ to $\xi_+$ can be seen in Figure \ref{ye}. 
$\xi_-$ is the length scale at which the density of paired 
vortices becomes small, while $\xi_+$ is the length scale at 
which the density of free vortices is large. In the case of 
the 2DYG, both $\xi_+$ and $\xi_-$ can be defined at all temperatures.

Defining the correlation length as the quantity $\xi_\pm$  
introduced by Simkin and Kosterlitz\cite{simkin97} is 
particularly important in the
case of the 2DYG. Because  $m$ is a relevant RG parameter, the
condition for terminating the integration of the recursion
relations will always be determined by $m$ becoming too large. This
will result in a featureless $\xi$ for $T<T_{KTB}$. But  
$\xi_-$, computed in accordance with Ref.~\onlinecite{simkin97} by 
using the condition that $y$ reaches an arbitrary low value, has a
clear physical meaning. 

\begin{figure}
\centerline{
\epsfig{file=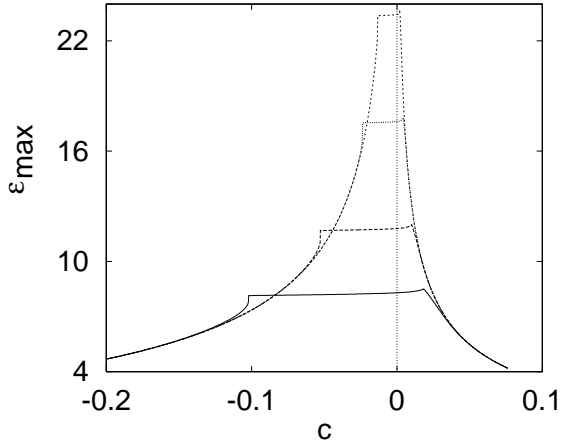,width=3.4in}}
\caption{$\epsilon_{max}$ (the logarithm of the correlation 
length) versus the integration constant $c\propto T/T_{KTB}-1$ for
$m_i=10^{-20}$, $10^{-15}$, $10^{-10}$, and $10^{-7}$. The plateau 
is where the correlation length exceeds $\lambda_\perp$.}
\label{corrlength}
\end{figure}

The correlation length, defined as explained above, is plotted in 
Figure \ref{corrlength}, for various initial values of $m$, versus 
$c\propto T-T_{KTB}$. As one can see, for each value of $m_i$, 
the correlation length starts to diverge near $T_{KTB}$ but then 
stops growing and reaches a plateau. $\xi$ to the left of the 
plateau is determined by $y$ being small: $y=0.01$. 
$\xi$ in the temperature range of the plateau is determined by 
the large $m$ cutoff, and $\xi$ to the right of the plateau is 
determined by the large $y$ cutoff. The abrupt changes in $\xi$ are 
due to the cutoff in the integration shifting one condition to
another. The cutoff value for the small value of $y$ is rather 
arbitrary but the same qualitative behavior is found using other 
values. Furthermore, as one can see from Figure \ref{ye}, $y=0.01$ 
does represent a value above which  the vast majority of the 
vortices are paired.  

Two temperature quantities can be studied by examining
the behavior of $\xi$
in Figure \ref{corrlength}. The first is the temperature width 
of the plateau, $\tau_{fs}$. As $m$ is increased, $\tau_{fs}$ also 
increases and its dependence on $m$ is well characterized by
\begin{equation}
\tau_{fs}\propto 1/\ln^2 m.
\label{width}
\end{equation}
It has been previously found that the shift in $T_c$ for 
a layered systems has this inverse logarithmic square 
dependence upon the strength of the interlayer 
coupling,\cite{pierson92,pierson95a,hikami80} indicating a deeper
reason for this dependence. The second quantity that can be studied,
at least qualitatively, is the transition temperature. Before the 
divergence of the correlation length is cut off, one can see that 
the effective transition temperature has shifted down. This is as 
expected since the vortex pairs interact more weakly. The temperature 
shift should have the same dependence as the quantity in 
Eq.~(\ref{width}) but this can not be verified here because the critical 
behavior is rounded off.

We now examine the renormalization of the stiffness constant
(superfluid density) $K$. For the 2DCG, this quantity, which is 
evaluated for $\epsilon=\infty$, first decreases linearly with 
increasing temperature and then decreases more rapidly near 
$T_{KTB}$ with the temperature dependence of the well-known 
square-root cusp. Precisely at $T_{KTB}$, it jumps to zero 
from its universal value of 4 (in these units). 

\begin{figure}
\centerline{
\epsfig{file=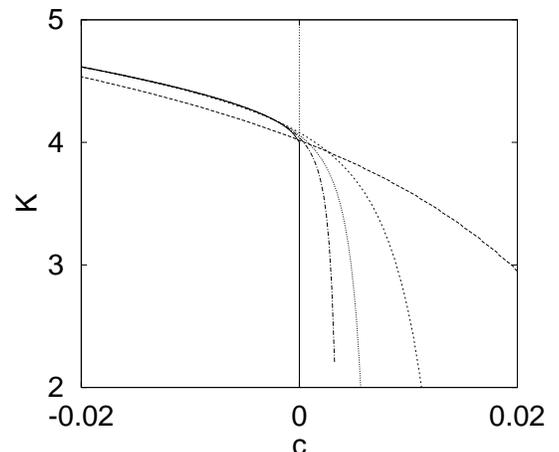,width=3.4in}}
\caption{The renormalized stiffness constant $K(\xi)$
as a function of temperature for $m_i=0$, $10^{-20}$, $10^{-15}$,
$10^{-10}$, and $10^{-5}$.}
\label{Ke}
\end{figure}

In the case of the 2DYG, $K(\infty)$ is equal to zero, independent of 
temperature, because there are free vortices at all temperatures. 
Hence, one must look at $K$ for length scales $\xi$, as shown in Figure 
\ref{corrlength}. $\xi$ is the pertinent length scale especially for 
$T\ll T_{KTB}$ where the cutoff determining $\xi$ is $y(\epsilon)$ 
small, because it characterizes vortex pairs, and it is the vortex 
pair properties (in the presence of a current) that allow one to 
measure $K$ via the $I$-$V$ exponent. $K(\xi)$ is plotted
in Fig.~\ref{Ke} for various initial values of $m$. One can see for 
$m_i=0$, that $K$ has the expected behavior for the 2DCG including the jump 
from universal value of 4 (for our definition
of this parameter) to zero. As the value of $m_i$ 
is increased, the jump becomes considerably more rounded. The temperature width 
over which the jump is smoothed is given by $\tau_{fs}$. At the largest value
of $m_i$, one can see that even the $T\ll T_{KTB}$ values of $K$ are depressed.

In this work, we have considered the effect of the {\it 
intrinsic} finite size scale $\lambda_\perp$, as opposed 
to the externally imposed finite size of the system. While 
their effects can be expected to be qualitatively 
similar and difficult to distinguish experimentally (see 
Ref.~\onlinecite{williams86} for effects of a finite system), 
there is an important difference. In both cases there are 
free vortices below the transition temperature. But the 
effect of the length $\lambda_\perp$ on the vortex 
interaction is a little larger than that of the system size. 
As a result, we expect that the transition temperature will
not be depressed as much in the finite extent system. There 
may be more differences in terms of the dynamics, and this 
deserves further study.

To close this section we remark on the implications of this work for
the 2D massive SG Hamiltonian itself. The recursion relations found here 
for the 2D massive sine-Gordon
differ slightly from those of Ref.~\onlinecite{ichinose94}. As discussed
in Section \ref{sec:rg}, the sign of the correction to the recursion
relation for $K$ found here is opposite to that found by those 
authors.\cite{ichinose94} We nonetheless feel that the behavior claimed
by those authors would be affected only quantitatively because, as noted 
above, it is the $m$ correction to the $y$ recursion relation that 
determines the effect of the mass on the system.

\section{Experimental Ramifications and the effect of currents}    
\label{sec:exp}    
In this paper, we have rigorously verified the well-known notion 
that the intrinsic finite size effect characterized by the relevant
parameter $\lambda_\perp$ should eventually dominate
any critical behavior. One should therefore ask why this effect is not
more prevalent in experimental studies. To address this question, one
must incorporate the effect of an applied current since most
experimental studies of KTB behavior in superconductors involve such 
a current. There one expects to see finite size effects for $T<T_{KTB}$ 
when the probing length of the current $r_c$ ($\propto 1/I)$ exceeds 
$\lambda_\perp$.\cite{kadin83,pierson99} We know from the work here 
that the renormalized value of $\lambda_\perp$ decreases nearly 
exponentially, [see Eq.~(\ref{rrc}),] and so one would expect 
finite size effects to always be seen. However, through a deeper 
examination of this condition, we will see that this is not the 
case. To do that, we must know the behavior of $r_c$ under 
renormalization.

Sujani {\it et al.}\cite{shenoy94} have derived the recursion 
relations for the 2DCG in the presence of a current. These can also be 
obtained by taking the 2D limit of the recursion relations of the 
layered vortex gas in the presence of a current, as derived in 
Ref.~\onlinecite{pierson95b}. (See also the work of 
Cserti\cite{cserti} on 2D dislocation systems and the effect of stress.) 
The latter results include a correction to the recursion relation for 
$K$ due to the current $I$ not found in the former. 
Otherwise, the results are in agreement.\cite{rrdetail}

For the case of the 2DYG in the presence of a current, the recursion 
relations would consist of those for the 2DCG along with the first 
order corrections from both the finite size variable $m$ calculated 
here and the current calculated in Refs.~\protect\onlinecite{shenoy94} 
and Refs.~\protect\onlinecite{pierson95b}. The relevant recursion 
relation here is that for $I$: $dI/d\epsilon=I$.

Let us now return to the condition for seeing inherent finite size 
effects in the presence of a current: $r_c\gg\lambda_\perp$. Even 
though $\lambda_\perp$ decreases nearly exponentially under 
renormalization, it will never get smaller than $r_c$ unless 
as its bare value starts out smaller than the bare value of 
$r_c$, because $r_c$ is decreasing exponentially under 
renormalization. This explains the experimental situation.

We now briefly address the expected effect of finite size
effects on the current-voltage relationship for the 2DCG. 
The first order effects have been worked out for the 
layered case.\cite{pierson97} Taking the 2D limit of 
Equation (11) in Ref.~\onlinecite{pierson97}, (i.e., $I_c^1
\rightarrow 0$,) one obtains\cite{ivnote} 
\begin{equation}
V=I(a+bI^{2\alpha})^{1/2},
\label{ivfinsiz}
\end{equation}
where $\alpha$ is the familiar 2DCG $I$-$V$ exponent and 
$a$ and $b$ are temperature dependent parameters. As 
expected,\cite{simkin97,repaci96} the
$I$-$V$ relationship is ohmic at small currents, due to 
free vortices induced by finite size effects, and then
becomes a power-law due to vortex pairs for larger currents. 
The crossover from ohmic to non-ohmic behavior is determined by 
the vortex correlation length $\xi_-$, thereby setting
limits on $a$ and $b$. Eq.~(\ref{ivfinsiz})
is based on approximations assuming weak current. We 
expect that a more rigorous treatment of the $I$-$V$ relation 
in the presence of finite size effects would include extensions 
to finite current, such as a dynamic\cite{ammirata99,pierson99} 
or finite size scaling approach.

\section{Summary}    
\label{sec:summ} 

In this work, we have derived the recursion relations for the 
two-dimensional Yukawa gas, relevant to vortices in 2D 
superconductors. We have examined the renormalized stiffness 
constant as a function of length scale, the RG flows, and the 
width of the region around $T_{KTB}$ that is no longer critical 
because of the finite size effect. We have also examined the 
condition for observing finite size effect in electrical 
transport measurements on superconducting films. 

\acknowledgements    
The authors gratefully acknowledge useful conversations with 
Sergey Simanovsky and Hocine Bahlouli. Acknowlegement is made 
by SWP to the donors of The Petroleum 
Research Fund, administered by the ACS, for support of this research.
SWP also expresses his gratitude to the Theoretical Physics Institute 
at the University of Minnesota for their hospitality while part of this
work was carried out.


\begin{references}
\bibitem{rrdetail}While not at first apparent, the recursion relations 
for the electric current from Refs.~\protect\onlinecite{shenoy94} and 
Refs.~\protect\onlinecite{pierson95b} are the same. The key is that 
the electric field variable that produces the current in 
Ref.~\protect\onlinecite{shenoy94} is a product of the current
and the stiffness constant.
\end{references}
\end{document}